\documentclass[12pt,superscriptaddress,showkeys,showpacs,] {revtex4}
\usepackage{graphicx}
\begin{document}

\title[Short Title]{Three-party quantum secure direct
 communication based on Greenberger-Horne-Zeilinger states}
\author{Xing-Ri Jin}
\author{Xin Ji}
\author{Ying-Qiao Zhang}
\author{Shou Zhang\footnote{E-mail: szhang@ybu.edu.cn} }
\affiliation{Department of Physics, College of Science, Yanbian
University, Yanji, 133002, PR China}
\author{Suc-Kyoung Hong}
\author{Kyu-Hwang Yeon}
\affiliation{Department of Physics,  Institute for Basic  Science
Research, College of Natural Science, Chungbuk National University,
Cheonju, Chungbuk 361-763, South Korea}

\author{Chung-In Um}
\affiliation{Department of Physics, College of Science, Korea
University, Seoul 136-701, South Korea}

\begin{abstract} We present a three-party simultaneous quantum secure direct
communication(QSDC)
 scheme by using Greenberger-Horne-Zeilinger(GHZ) states.
 This
scheme can be directly generalized to $N$-party QSDC by using
$n$-particle GHZ states. We show that the many-party simultaneous
QSDC scheme is secure not only against the intercept-and-resend
attack but also against the disturbance attack. 
 \end{abstract}
 \maketitle
 Quantum key distribution(QKD) is an ingenious
application of quantum mechanics, in which two remote legitimate
users(Alice and Bob) establish a shared secret key through the
transmission of quantum signals and use this key to
encrypt(decrypt) the secret messages. Since Bennett and Brassard
presented the pioneer QKD protocol in 1984\cite{BB84}, a lot of
QKD protocols have been advanced\cite{BBM92, BS92, B98, CA00}. On
the other hand, a novel concept, QSDC has been
proposed\cite{AB02,KB02,FGD03,MAN0518,QYC03, TC04, DENG04}.
Different from QKD whose object is to distribute a common key
between the two remote legitimate users, QSDC can transmit the
secret messages directly without creating a key to encrypt them
beforehand.

Recently, Beige \emph{et al}\cite{AB02} presented a QSDC scheme
based on single photon. Bostr$\rm{\ddot{o}}$m and
Felbinger\cite{KB02} put forward a ping-pong QSDC scheme by using
Einstein-Podolsky-Rosen(EPR) pairs and Deng \emph{et
al}\cite{DENG04} proposed a QSDC scheme based on single photon
four states. In their schemes, QSDC is only one-way communication.
Based on the idea of a ping-pong QSDC scheme, Nguyen\cite{BAN04}
proposed a quantum dialogue scheme (the quantum dialogue is
actually two-way communication) by using EPR pairs. However, an
eavesdropper who adopts the intercept-and-resend attack strategy
can steal the secret messages without being detected. More
recently, Gao \emph{et al}\cite{GAO05} presented a simultaneous
QSDC scheme between the central party and other many parties based
on entanglement swapping. The scheme shows how the many parties
transmit secret message to one party.

In this paper, we propose a three-party(Alice, Bob and Charlie)
simultaneous QSDC scheme by using three-particle GHZ states. In
our scheme, Alice can obtain the secret messages of Bob and
Charlie. Also, Bob(or Charlie) can obtain the secret messages of
Alice and Charlie(or Bob). Their secret messages exchange is
secure and simultaneous. Indeed, this scheme can be directly
generalized to $N$-party QSDC by using $n$-particle GHZ states.

Now we propose the three-party simultaneous QSDC scheme. First, we
write the eight GHZ entangled states in two different bases as
follows:
\begin{eqnarray}\label{e1}
&|\psi_{000}\rangle_{abc}&=\frac{1}{\sqrt{2}}(|000\rangle+|111\rangle)_{abc}
\cr\cr&&=\frac{1}{2}[|+\rangle_a(|+\rangle_b|+\rangle_c+
|-\rangle_b|-\rangle_c)+|-\rangle_a(|+\rangle_b|-\rangle_c+|-\rangle_b|+\rangle_c)],
\end{eqnarray}
\begin{eqnarray}\label{e2}
&|\psi_{001}\rangle_{abc}&=\frac{1}{\sqrt{2}}(|000\rangle-|111\rangle)_{abc}
\cr\cr&&=\frac{1}{2}[|+\rangle_a(|+\rangle_b|-\rangle_c+
|-\rangle_b|+\rangle_c)+|-\rangle_a(|-\rangle_b|-\rangle_c+|+\rangle_b|+\rangle_c)],
\end{eqnarray}
\begin{eqnarray}\label{e3}
&|\psi_{010}\rangle_{abc}&=\frac{1}{\sqrt{2}}(|100\rangle+|011\rangle)_{abc}\cr\cr&&
=\frac{1}{2}[|+\rangle_a(|+\rangle_b|+\rangle_c+
|-\rangle_b|-\rangle_c)-|-\rangle_a(|+\rangle_b|-\rangle_c+|-\rangle_b|+\rangle_c)],
\end{eqnarray}
\begin{eqnarray}\label{e4}
&|\psi_{011}\rangle_{abc}&=\frac{1}{\sqrt{2}}(|100\rangle-|011\rangle)_{abc}\cr\cr&&=\frac{1}{2}
[|+\rangle_a(|+\rangle_b|-\rangle_c+
|-\rangle_b|+\rangle_c)-|-\rangle_a(|+\rangle_b|+\rangle_c+|-\rangle_b|-\rangle_c)],
\end{eqnarray}
\begin{eqnarray}\label{e5}
&|\psi_{100}\rangle_{abc}&=\frac{1}{\sqrt{2}}(|010\rangle+|101\rangle)_{abc}\cr\cr&&=\frac{1}{2}
[|+\rangle_a(|+\rangle_b|+\rangle_c-
|-\rangle_b|-\rangle_c)+|-\rangle_a(|+\rangle_b|-\rangle_c-|-\rangle_b|+\rangle_c)],
\end{eqnarray}
\begin{eqnarray}\label{e6}
&|\psi_{101}\rangle_{abc}&=\frac{1}{\sqrt{2}}(|010\rangle-|101\rangle)_{abc}\cr\cr&&=\frac{1}{2}
[|+\rangle_a(|+\rangle_b|-\rangle_c-
|-\rangle_b|+\rangle_c)+|-\rangle_a(|+\rangle_b|+\rangle_c-|-\rangle_b|-\rangle_c)],
\end{eqnarray}
\begin{eqnarray}\label{e7}
&|\psi_{110}\rangle_{abc}&=\frac{1}{\sqrt{2}}(|110\rangle+|001\rangle)_{abc}\cr\cr&&=\frac{1}{2}
[|+\rangle_a(|+\rangle_b|+\rangle_c-
|-\rangle_b|-\rangle_c)+|-\rangle_a(|-\rangle_b|+\rangle_c-|+\rangle_b|-\rangle_c)],
\end{eqnarray}
\begin{eqnarray}\label{e8}
&|\psi_{111}\rangle_{abc}&=\frac{1}{\sqrt{2}}(|110\rangle-|001\rangle)_{abc}\cr\cr&&=\frac{1}{2}
[|+\rangle_a(|+\rangle_b|-\rangle_c-
|-\rangle_b|+\rangle_c)+|-\rangle_a(|-\rangle_b|-\rangle_c-|+\rangle_b|+\rangle_c)],
\end{eqnarray}
where
\begin{equation}\label{e9}
|+\rangle=\frac{1}{\sqrt{2}}\ (|0\rangle+|1\rangle),
\end{equation}
\begin{equation}\label{e10}
|-\rangle=\frac{1}{\sqrt{2}}\ (|0\rangle-|1\rangle).
\end{equation}

Alice, Bob and Charlie agree on that Alice can perform the four
unitary operations
\begin{eqnarray}\label{e11}
&& I=|0\rangle\langle0|+|1\rangle\langle1|,\quad\quad
\sigma_{x}=|0\rangle\langle1|+|1\rangle\langle0|, \cr\cr
&&i\sigma_{y}=|0\rangle\langle1|-|1\rangle\langle0|,\quad\quad
\sigma_{z}=|0\rangle\langle0|-|1\rangle\langle1|,
\end{eqnarray}
and encode two bits classical information as
\begin{eqnarray}\label{e12}
 I\rightarrow 00,\quad\quad
\sigma_{x}\rightarrow 01,\quad\quad i\sigma_{y}\rightarrow
10,\quad\quad \sigma_{z}\rightarrow 11.
\end{eqnarray}
Bob and Charlie can only perform the two unitary operations
\begin{eqnarray}\label{e13}
I=|0\rangle\langle0|+|1\rangle\langle1|,\quad\quad
i\sigma_{y}=|0\rangle\langle1|-|1\rangle\langle0|,
\end{eqnarray}
respectively, and encode one bit classical information as
$I\rightarrow 0$, $i\sigma_{y}\rightarrow 1$.

 The three-party
simultaneous QSDC scheme can be achieved with three steps:

 (Step i.)
Suppose Alice, Bob and Charlie want to exchange their messages
simultaneously. At first, Alice prepares a set of $N$ groups
three-particle GHZ states randomly in one of the eight
three-particle GHZ states$(|\psi_{ijk}\rangle_{abc},i,j,k=0,1)$.
Then Alice sends $N$ groups $b$ particles to Bob and  $c$
particles to Charlie, respectively. Here, since Bob and Charlie
can not distinguish the particles $a$, $b$ and $c$, Alice must let
them know which one they have received.

(Step ii.) Bob and Charlie confirm Alice that they have received
all the particles $b$ and $c$, respectively. Then Bob(anyone of
the two parties Bob and Charlie, we say, Bob) selects randomly a
sufficiently large subset of particles from the $N$ groups $b$
particles, which we call the $M$ groups $b$ particles and measures
each of them using one of the two measuring
bases$(|0\rangle,|1\rangle$) or ($|+\rangle,|-\rangle )$ randomly.
And Bob tells Charlie and Alice the position, the measuring basis
and the measurement result for each of the $M$ groups $b$
particles via classical channel. Then Alice and Charlie measure on
the corresponding $M$ groups $a$ particles and $M$ groups $c$
particles using the same measuring bases, respectively. Then
Charlie tells Alice the measurement result for each of the $M$
groups $c$ particles. According to the measurement results of Bob,
 Charlie by herself, Alice can determine,
 through the error rate, whether there is any eavesdropping in the channel. If
 the error rate is high, Alice concludes that the channel is not
 secure, and halts the communication. Otherwise, Alice, Bob
 and Charlie continue to the next step.

(Step iii.) The particles leftover are called the $K$
groups($K=N-M$) after checking the eavesdropping. After
determining the security of quantum channel, Bob and Charlie
encode each of the $K$ groups $b$ particles and $c$ particles with
one of the two unitary operations $I$ and $i\sigma_{y}$,
respectively, according to their secret messages. Then they return
the $K$ groups $b$ particles and $c$ particles to Alice, and Alice
encodes each of the $K$ groups $a$ particles with one of the four
unitary operations $I$, $\sigma_{x}$, $i\sigma_{y}$ and
$\sigma_{z}$ according to her secret message. Then she performs a
three-particle GHZ-basis measurement on $K$ groups $a$, $b$ and
$c$ particles and publicly announces her measurement result and
initial three-particle GHZ states. According to her measurement
result, initial three-particle GHZ states and the unitary
operation performed by herself, Alice can read out the secret
messages of Bob and Charlie. Also, Bob(or Charlie) can read out
the secret messages of Alice and Charlie(or Bob) according to
Alice's measurement result, initial three-particle GHZ states and
the unitary operation performed by themselves.

For example, if Alice initially prepares a GHZ state in
$|\psi_{000}\rangle_{abc}$, she performs the unitary operation
$\sigma_{x}$ on particle $a$. Bob and Charlie perform the unitary
operation $I$ and $i\sigma_{y}$ on particles $b$ and $c$,
respectively. So the $|\psi_{000}\rangle_{abc}$ becomes
$|\psi_{101}\rangle_{abc}$, namely, $\sigma_{x}^{a}\otimes I^{b}
\otimes
i\sigma_{y}^{c}|\psi_{000}\rangle_{abc}=|\psi_{101}\rangle_{abc}$.
According to the unitary operation $\sigma_{x}$ performed by
herself and the measurement result $|\psi_{100}\rangle_{abc}$ of
three-particle GHZ state, Alice can read out Bob's and Charlie's
secret messages $0$ and $1$, respectively. Similarly, Bob(or
Charlie) can also read out Alice's and Charlie's(or Bob's) secret
messages $01$ and $1$(or $0$), respectively. So the three-party
simultaneous QSDC has been successfully completed.

Then we continue to discuss how the current scheme resist the
intercept-and-resend attack and disturbance attack, respectively.

(1). The intercept-and-resend attack: We suppose that
Eve(eavesdropper) prepares some the single particles $B$ and $C$
in the state ($|0\rangle$, $|1\rangle$) or ($|+\rangle$,
$|-\rangle$) randomly. When Alice sends particles $b$ and $c$ to
Bob and Charlie, Eve intercepts the particles $b$ and $c$ and
keeps them with her, and sends the particles $B$ and $C$ to Bob
and Charlie, respectively. Bob and Charlie, will take $B$ and $C$
for $b$ and $c$, encode their secret messages on particles $B$ and
$C$ and send them back to Alice. Eve intercepts the encoded
 particles $B$ and $C$, and performs single particles measurements
 on them. Since the
 particles $B$ and $C$ are prepared by Eve,  Eve can obtain Bob's and
 Charlie's encoding messages according to the original states of particles $B$ and
 $C$ and the measurement results performed by herself.
 Then Eve encodes the same messages on the particles $b$ and
 $c$ and sends them back to Alice. According to Alice's measurement result, initial
three-particle GHZ states, Eve can read out Alice's secret
messages. Clearly, Alice, Bob and Charlie not only exchange their
messages simultaneously but also leak to Eve. This is the
intercept-and-resend attack. However, the method described in Step
ii. can resist the attack, namely, Alice, Bob and Charlie
collaborate to select randomly a sufficiently large subset of
particles from the $N$ groups to check whether there is any
eavesdropping in the channel through analyzing the error rate. If
there are no attacks, the measurement result of Alice, Bob and
Charlie should have deterministic correlation according to
Eqs.(1)-(8). For example, we suppose that the initial GHZ state is
prepared in $|\psi_{000}\rangle_{abc}$, if there is no Eve in the
line, the measurement result of Alice, Bob and Charlie is in
$|000\rangle$ or
$|111\rangle$($|+++\rangle,|+--\rangle,|-+-\rangle$ or
$|--+\rangle$). If Eve intercepts the particles $b$ and $c$ and
keeps them with her, and sends the particles $B$ and $C$ to Bob
and Charlie. If the particles $B$ and $C$ are prepared in states
$|0\rangle_{B}$ and $|1\rangle_{C}$, respectively.  Alice, Bob and
Charlie select the measuring bases $|0\rangle,|1\rangle$
 $(|+\rangle,|-\rangle )$, their the
measurement result is $|001\rangle_{aBC}$ or
$|101\rangle_{aBC}$($|opq\rangle_{aBC},o,p,q=+,-$). According to
Eq.(1) Eve's eavesdropping will be detected, because her
eavesdropping introduces a error rate with 1(1/2). If the
particles $B$ and $C$ are prepared in state $|0\rangle_{B}$ and
$|0\rangle_{C}$ or $|1\rangle_{B}$ and $|1\rangle_{C}$,
respectively, Eve's eavesdropping introduces a error rate with
1/2(1/2). Thus our scheme is secure against the
intercept-and-resend attack. In Nguyen's quantum dialogue
scheme\cite{BAN04}, the secret information can be completely
leaked to an eavesdropper who adopts the intercept-and-resend
attack strategy by using EPR pairs without being detected at all.

(2). Disturbance attack: Eve can intercept the particles $b$ and
 $c$ when the particles are transmitted from Bob and Charlie to
 Alice.
Eve may either measure on the particles $b$ and
 $c$\cite{QYC03} or perform one of the unitary operations $I$ and $i\sigma_{y}$ on them.
 By doing so, the entanglement between particles $b$ and
 $c$ is destroyed or the phase of the entanglement is changed. In this case, Eve can only remain undetected and
 the information encoded by Bob and Charlie are nothing but a
 random sequence of bits that contains no information.
To resist the disturbance attack, Bob and Charlie collaborate to
announce publicly the positions of their particles and a part of
their secret messages to Alice to check whether the particles
travelling from Bob's and Charlie's sites to Alice's site have
been attacked. If the particles are attacked, the eavesdropper Eve
can not get any useful information but interrupt the
transmissions.

Therefore, our three-party simultaneous QSDC scheme is secure not
only against the intercept-and-resend attack but also against
disturbance attack. Our scheme improves Nguyen's quantum dialogue
scheme\cite{BAN04} to realize a three-party simultaneous QSDC.
Comparing with Gao's scheme\cite{GAO05} in which only two
parties(Alice and Bob) transmit the secret massages to one
party(Charlie) based on entanglement swapping, respectively, our
scheme makes the three parties(Alice, Bob and Charlie)
simultaneously exchange their secret massages.

Now let us generalize the three-party simultaneous QSDC scheme to
$N$-party case. Alice, Bob, Charlie, ..., and Zach agree on that
Alice can perform one of the four unitary operantions($I,
\sigma_{x}, i\sigma_{y}$ and $\sigma_{z}$) and Bob, Charlie, ...,
and Zach can perform one of the two unitary operations($I$ and
$i\sigma_{y}$). The $N$-party simultaneous QSDC scheme can be
achieved with three steps:

(Step I.) Suppose Alice, Bob, Charlie, ...,  and Zach want to
exchange their messages simultaneously. Alice prepares a set of
$N$ groups $n$-particle GHZ states randomly in one of the $2^{n}$
$n$-particle GHZ
states$(|\psi_{ijk...y}\rangle_{abc....z},i,j,k,...,y=0,1)$. Then
Alice sends $N$ groups
 $b$ particles to Bob, $c$ particles to Charlie, ..., $z$ particles to Zach.
 Here, since Bob,
Charlie, ..., and Zach can not distinguish the particles $a$, $b$,
$c$, ... and $z$, Alice must let them know which particles they
have received.

(Step II.) Bob, Charlie, ..., and Zach confirm Alice that they
have received all the particles $b$, particles $c$, ..., and
particles $z$, respectively. Then Bob(anyone of the $N-1$ parties
Bob, Charlie, ..., Zach, we say, Bob) selects randomly a
sufficiently large subset of particles from the $N$ groups $b$
particles, which we call the $M$ groups $b$ particles, and
measures each of them using one of the two measuring
bases$(|0\rangle,|1\rangle$) or ($|+\rangle,|-\rangle )$ randomly.
Bob tells Alice, Charlie, ..., and Zach the position, the
measuring basis and the measurement result for each of the $M$
groups $b$ particles via classical channel. Then Alice, Charlie,
..., and Zach measure $M$ groups $a$ particles, $c$ particles,
..., $z$ particles using the same measuring bases, respectively.
Then they tell Alice their measurement results for each of the
particles. According to the measurement results of Bob, Charlie,
..., Zach and herself, Alice can determine whether there is any
eavesdropping. If the error rate is high, Alice concludes that the
channel is not secure, and halts the communication. Otherwise,
Alice, Bob, Charlie, ..., and Zach continue the the next step.

(Step III.) The particles leftover are called the $K$ groups
particles($K=N-M$) after the checking eavesdropping. After the
security  checking of the quantum channel, Bob, Charlie, ..., and
Zach encode each of the $K$ groups $b$ particles, $c$ particles,
..., $z$ particles with one of the two unitary operations $I$ and
$i\sigma_{y}$, respectively, according to their secret messages.
After receiving the $K$ groups $b$ particles, $c$ particles, ...,
and $z$ particles, Alice encodes each of the $K$ groups $a$
particles with one of the four unitary operations $I$,
$\sigma_{x}$, $i\sigma_{y}$ and $\sigma_{z}$ according to her
secret message. Then Alice performs a $n$-particle GHZ-basis
measurement on $K$ groups $a$, $b$, $c$, ..., and $z$ particles
and publicly announces her measurement result and initial
$n$-particle GHZ states. According to her measurement result,
initial $n$-particle GHZ state and the unitary operation performed
by herself, Alice can read out the secret messages of Bob,
Charlie, ..., and Zach. Also, Bob(Charlie, ..., Zach) can read out
the secret message of the others. So the $N$-party simultaneous
QSDC has been successfully completed.

In conclusion, we have proposed a three-party simultaneous QSDC
scheme by using three-particle GHZ states. We also generalize this
scheme to $N$-party simultaneous QSDC by using $n$-particle GHZ
states. In our scheme an eavesdropper Eve's action can be detected
efficiently. Therefore, the simultaneous QSDC scheme is secure not
only against the intercept-and-resend attack but also against
disturbance attack.


\begin{thebibliography}{999}
\bibitem{BB84} C. H. Bennett, G. Brassard, Proc. IEEE Int.
Conf. on Computers, Systems and Signal Processing, Bangalore,
India, (IEEE, New York), P (1984) 175.
\bibitem{BBM92} C. H. Bennett, G. Brassard, N. D. Mermin, Phys.
Rev. Lett. 68 (1992) 557.
\bibitem{BS92} C. H. Bennett, S. J. Wiesner, Phys.
Rev. Lett. 69 (1992) 2881.
\bibitem{B98} D. Bruss, Phys. Rev. Lett.  81 (1998) 3018.
\bibitem{CA00} A. Cabell, Phys. Rev. Lett. 85 (2000) 5635.
\bibitem{AB02} A. Beige, B. G. Engler, C. Kurtsiefer, H.
Weinfurter, Acta Phys. Pol. A 101 (2002) 357.
\bibitem{KB02} K. Bostr\"{o}m,  T. Felbinger, Phys. Rev.
Lett. 89 (2002) 187902.
\bibitem{FGD03} F. G. Deng, G. L. Long, X. S. Liu, Phys. Rev. A 68
(2003) 042317.
\bibitem{MAN0518} Z. X. Man, Z. J. Zhang,  Y. Li, Chin. Phys.
Lett. 22 (2005) 18.
\bibitem{QYC03} Q. Y. Cai, Phys. Rev. Lett. 91 (2003) 109801.
\bibitem{TC04} T. Gao, F. L. Yan,  Z. X. Wang, Nuovo Cimento B
119 (2004) 313.
\bibitem{DENG04} F. G. Deng, G. L. Long, Phys. Rev. A 69 (2004)
052319.
\bibitem{BAN04} B. A. Nguyen, Phys. Lett. A 328
(2004) 6.
\bibitem{GAO05} T. Gao, F. L. Yan, Z. X. Wang, J. Phys. A 38 (2005)
5761.

\end{thebibliography}
\end{document}